\documentstyle[galley,epsf]{mn}
\newif\ifAMStwofonts

\ifoldfss
  \ifCUPmtlplainloaded \else
    \NewTextAlphabet{textbfit} {cmbxti10} {}
    \NewTextAlphabet{textbfss} {cmssbx10} {}
    \NewMathAlphabet{mathbfit} {cmbxti10} {} 
    \NewMathAlphabet{mathbfss} {cmssbx10} {} 
  \fi
  \ifAMStwofonts
    \ifCUPmtlplainloaded \else
      \NewSymbolFont{upmath} {eurm10}
      \NewSymbolFont{AMSa} {msam10}
      \NewMathSymbol{\upi}     {0}{upmath}{19}
      \NewMathSymbol{\umu}     {0}{upmath}{16}
      \NewMathSymbol{\upartial}{0}{upmath}{40}
      \NewMathSymbol{\leqslant}{3}{AMSa}{36}
      \NewMathSymbol{\geqslant}{3}{AMSa}{3E}

      \let\leq=\leqslant 
      \let\geq=\geqslant 
    \fi
  \fi
\fi 

\ifnfssone
  \newmathalphabet{\mathit}
  \addtoversion{normal}{\mathit}{cmr}{m}{it}
  \addtoversion{bold}{\mathit}{cmr}{bx}{it}
  \newmathalphabet{\mathbfit} 
  \addtoversion{normal}{\mathbfit}{cmr}{bx}{it}
  \addtoversion{bold}{\mathbfit}{cmr}{bx}{it}
  \newmathalphabet{\mathbfss} 
  \addtoversion{normal}{\mathbfss}{cmss}{bx}{n}
  \addtoversion{bold}{\mathbfss}{cmss}{bx}{n}
  \ifAMStwofonts
    \ifCUPmtlplainloaded \else
      \UseAMStwoboldmath
      \makeatletter
      \new@mathgroup\upmath@group
      \define@mathgroup\mv@normal\upmath@group{eur}{m}{n}
      \define@mathgroup\mv@bold\upmath@group{eur}{b}{n}
      \edef\UPM{\hexnumber\upmath@group}
      \new@mathgroup\amsa@group
      \define@mathgroup\mv@normal\amsa@group{msa}{m}{n}
      \define@mathgroup\mv@bold\amsa@group{msa}{m}{n}
      \edef\AMSa{\hexnumber\amsa@group}
      \makeatother
      \mathchardef\upi="0\UPM19
      \mathchardef\umu="0\UPM16
      \mathchardef\upartial="0\UPM40
      \mathchardef\leqslant="3\AMSa36
      \mathchardef\geqslant="3\AMSa3E

      \let\leq=\leqslant 
      \let\geq=\geqslant 
    \fi
  \fi
\fi 

\ifnfsstwo
  \DeclareMathAlphabet{\mathbfit}{OT1}{cmr}{bx}{it}
  \SetMathAlphabet\mathbfit{bold}{OT1}{cmr}{bx}{it}
  \DeclareMathAlphabet{\mathbfss}{OT1}{cmss}{bx}{n}
  \SetMathAlphabet\mathbfss{bold}{OT1}{cmss}{bx}{n}
  \ifAMStwofonts
    \ifCUPmtlplainloaded \else
      \DeclareSymbolFont{UPM}{U}{eur}{m}{n}
      \SetSymbolFont{UPM}{bold}{U}{eur}{b}{n}
      \DeclareSymbolFont{AMSa}{U}{msa}{m}{n}
      \DeclareMathSymbol{\upi}{0}{UPM}{"19}
      \DeclareMathSymbol{\umu}{0}{UPM}{"16}
      \DeclareMathSymbol{\upartial}{0}{UPM}{"40}
      \DeclareMathSymbol{\leqslant}{3}{AMSa}{"36}
      \DeclareMathSymbol{\geqslant}{3}{AMSa}{"3E}

      \let\leq=\leqslant 
      \let\geq=\geqslant 
    \fi
  \fi
\fi 

\ifCUPmtlplainloaded \else
  \ifAMStwofonts \else 
    \def\upi{\pi}
    \def\umu{\mu}
    \def\upartial{\partial}
  \fi
\fi

\begin{document}
\title{High Resolution Spectroscopy of the high galactic latitude RV Tauri star CE Virginis}
\author[N. Kameswara Rao \and Bacham E. Reddy] 
       {N. Kameswara Rao, Bacham E. Reddy\\
       Indian Institute of Astrophysics, Bangalore 560034, India\\}
\date{Accepted .
      Received ;
      in original form  }

\pagerange{\pageref{firstpage}--\pageref{lastpage}}
\pubyear{}

\maketitle

\label{firstpage}

\begin{abstract}

Analysis of the surface composition of the suspected cool RV Tauri star CE Vir shows
no systematic trend in depletions of elements with respect to condensation temperature.
However, there is a significant depletion of the elements with respect to the first ionization
potential of the element. The derived Li abundance of log~$\epsilon$ (Li) = 1.5$\pm$0.2 indicates
production of Li in the star. Near infrared colours indicate sporadic 
dust formation close to the photosphere.
\end{abstract}


\section{Introduction}

RV Tauri stars are low mass late type supergiant pulsators that show deep and
shallow minima with periods ranging from 40 to 150 days. Their spectral types
range from F5 to K3 although most of them are earlier than G3 (Lloyd Evans 1999
). Preston et al. (1963) classified RV Tauris into three spectroscopic classes
RVA, RVB and RVC. The RVAs are normal oxygen rich group, and  RVBs show 
lines and bands of C\,{\sc I}, CH and CN where as  RVCs show weak metal lines and high
radial velocities, and are thought to be metal poor. The RVA and RVB may belong to disk
population. RVB and RVC stars are generally of earlier spectral type than RVA.
The evolutionary status of RV Tauri stars is thought to be post-AGB (Jura 1986, Alcolea
\& Bujarrabal 1991 ) or at the end of AGB phase (Gingold 1986). Photospheric
chemical composition is expected to give clues to their evolutionary status.
Recent abundance analyses of a sample of field RV Tauri stars (Giridhar, Rao \&
Lambert 1994; Gonzalez, Lambert \& Giridhar 1997a, 1997b; Giridhar, Lambert \& 
Gonzalez 1998, 2000) revealed  that stars with intrinsic metallicity [Fe/H]
$>$ $-$1.0, as indicated by S and Zn, possess abnormal surface abundance pattern 
that is quite distinctly different from either disk or halo stars. The elements
that have their condensation temperatures ($T_{\rm c }$ ) $>$ 1300 K are 
selectively depleted in such a way that the higher the $T_{\rm c }$  
higher is the depletion (relative to solar abundances). The condensation 
temperature adopted of a given element is the temperature at which half the atoms in 
gaseous environment condense out of the gas phase (Lodders 2003). 
It was suggested that the atmosphere accreted leftover gas from the site of
dust formation where several elements with higher $T_{\rm c }$ get lockedup
in grains. However it became clear that stars with intrinsic metallicities
($[Fe/H]$ ) $<$ $-$1 are not subjected to this  composition anomalies resulting 
from dust-gas (DG) separation and subsequent accretion of the winnowed gas (
Giridhar, Lambert \& Gonzalez 2000). It is not presently clear where the
site of dust formation and the dust-gas separation is taking place, whether
in a circumbinary disk (Van Winkle et al 1999) or in the stellar wind (or
circumstellar region) although many of the RV Tauri stars that exhibit 
the abundance anomalies are binaries.

\begin{figure*}
\epsfxsize=16truecm
\epsffile{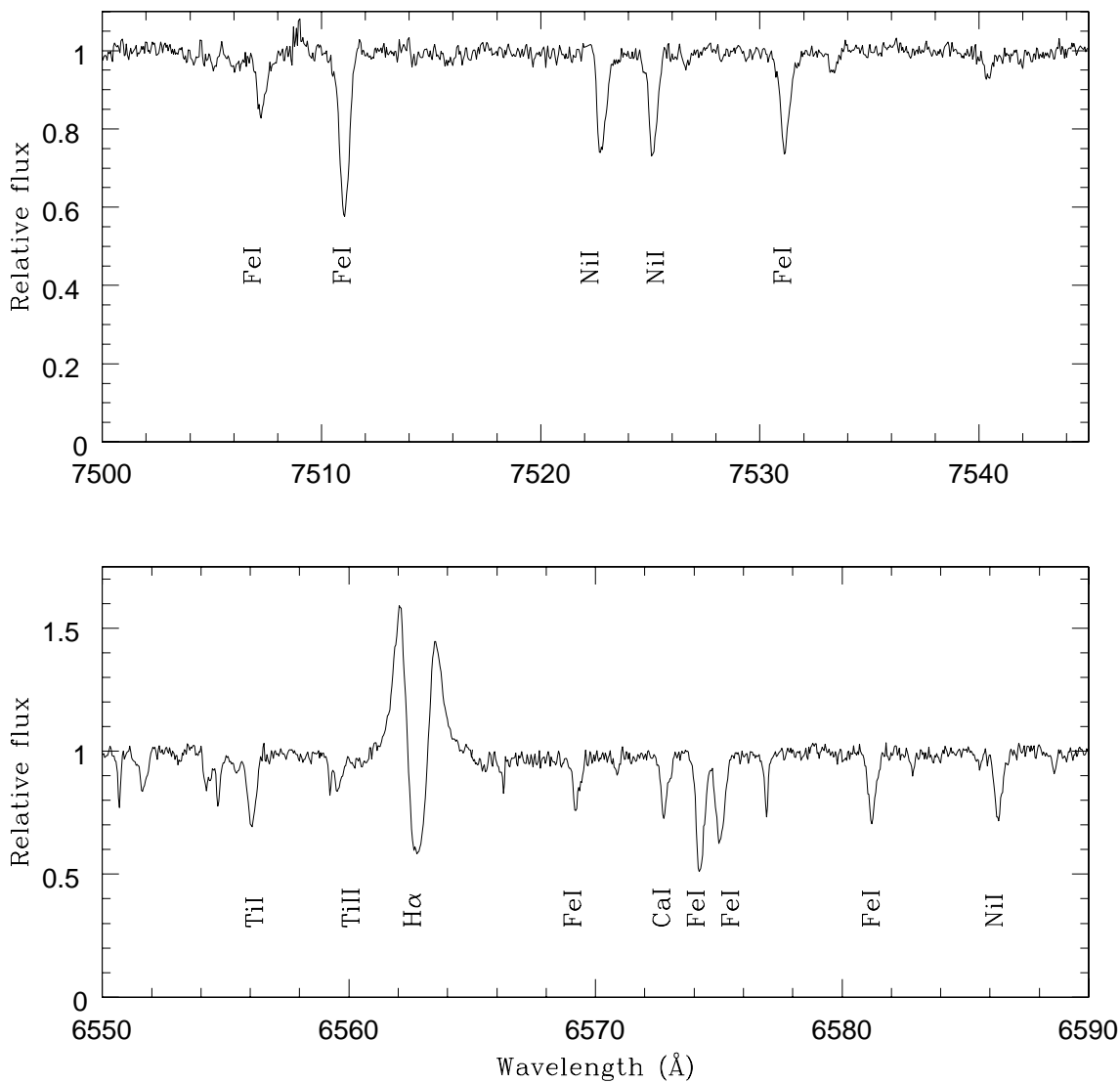}
\caption{Sample spectra of CE Vir showing $\lambda$7520~\AA\ and H$_{\alpha}$ regions.}
\end{figure*}

In reviewing the process of dust gas separation Giridhar, Lambert \& 
Gonzalez (2000) suggest that the absence of the anomalies in RVA stars is due
to either inefficient return of gas to the atmosphere and/or to the dilution
of this gas by the deeper convective envelope of the cooler stars. Since
RVAs are likely to turn into RVBs the initiation of the return of the winnowed
gas to the photosphere would be expected to happen in the RVA phase 
particularly in the coolest stars. Exploration of the coolest RVAs might give
some clues to how the process gets initiated. CE Virginis is one of the
coolest stars in which Giridhar et al (2000) find traces of abundance anomalies related
to winnowed gas -their DG index of 1. We undertook to explore this star 
spectroscopically in more detail. CE Vir is thought to be a SRd star but 
Gonzalez et al. (1997b: hereafter GLG) consider it as a cool member of RV Tauri class based
on the similarity with the bonified RV Tauri star DY Aql. CE Vir is also one of the
two stars in a sample of 21 RV Tau stars with a strong Li\,{\sc I} 6707~\AA\ line and also
located at high galactic latitude (57.8 degrees). Normally RV Tau stars at high galactic
latitudes are expected to be RVC stars with weak metal lines but Gonzalez et
al. (1997b) classify CE Vir as a RVA star. Moreover the analysis of Gonzalez et al.
is confined to a few elements only. Because of these interesting characteristics
we obtained high resolution spectra of CE Virginis.  

\section{Observations}

High-resolution optical spectra have been obtained with the Fiber-fed coude
echelle spectrometer (Rao et al. 2004) of the 2.3 meter Vainu Bappu Telescope (VBT) on
a few occassions. The spectra discussed and displayed here (Figures 1 to 4) have been obtained with VBT on 
17 and 18 February, 2004. The spectrometer operates in a littrow mode and 
consists of a six element collimater-camera system that operates both as a
collimater to the input beam and as a camera to the dispersed output beam. The
collimated beam passes through a cross disperser prism twice; once before
it reaches the echelle grating and once after the echelle grating. The
main dispersing element in the spectrometer is 
a 408 mm $\times$ 204 mm echelle grating of  52.6 gr/mm with
a blaze angle
of 70 degrees. The input beam size is 150~mm. The dispersed spectrum is 
recorded on either a 2~K $\times$ 4~K, 15 micron pixel CCD  or a 1~K X 1~K 24 micron pixel CCD 
camera. The star light is fed to the spectrometer from the prime focus of the
telescope by an optical fiber of 45 meter length. 

The present observations used 1~K $\times$ 1~K CCD system and covered the spectral region
of 4980 to 8050~\AA\ with gaps. The spectral resolving power achieved in these 
observation R = $\lambda/\Delta\lambda$
as estimated from $FWHM$ of weak terrestrial O$_{\rm 2}$ lines is 63,000. A signal to
noise of 70 was obtained on an average for each of the spectra. Image reductions
are performed using the $IRAF$ software.
Sample spectra are presented in Figure~1. 

The spectra at these phases looked free of any molecular lines except weak
CN lines at 8010~\AA\ region (which we tried to use for estimating N abundance -
discussed further in other sections). The only emission conspicuously present 
is H$\alpha$ (Figure~1) which has a central deep absorption flanked by emission
peaks on either side. Slight line asymmetries might be present to a few lines 
but no line doubling or splitting is evident at this resolution, thus the 
spectrum seemed to be aminable for abundance analysis. The spectrum appears 
similar to a 1996 July spectrum of CE Vir discussed by Gonzalez et al.
(1997b). The period of the light variations are very uncertain 
even though General Catalogue of Variable Stars (GCVS)  gives it as 67 days with uncertainty. Recent analysis of
the Hipparcos epoch photometry by Percy \& Kolin (2000) could not result in any
one definitive period. Thus the relative phase of Gonzalez et al.'s (1997b) 
observations to ours cannot be ascertained.
 
The star has a $T_{\rm eff}$ similar to Arcturus; as such we have used the Atlas of the
Arcturus spectrum (Hinkle et al 2000) for line identification along with
the spectrum of K2~III star DZAnd (Goswami, Rao \& Lambert 1998).  

\section{Radial Velocity}  
  
CE Virginis seems to be mildly variable in radial velocity. Although no 
systematic measurements are available, the star seem
to have been observed on three occasion's. Jones and Fisher (1984) obtained 
3 measurements with a mean of $-$70.4$\pm$0.7 km s$^{-1}$ at Mt.Stromlo; not
too different from the recent measurements. Table~1 lists the individual
measurements (both McDonald and VBT measured by us).

\begin{table}
\caption{Radial Velocity Measurements of CE Vir}
\begin{tabular}{lrl}
\hline
               &  JD   &  km s$^{-1}$     \\
\hline\hline\\
MtStromlo  & 2441380.2 &$-$70.5            \\
           & 1464.1 &$-$69.4            \\
           & 1466.0 &$-$71.2             \\
Mcdonald   & 2450182.72 &$-$73.1 $\pm1.2$    \\
           & 1286.65 &$-$77.0 $\pm1.1$    \\
VBT        & 2453053.45 &$-$69.8 $\pm1.1$    \\
\hline
\end{tabular}
\end{table}

Based on the light curve it is not clear whether CE Vir is a SRd variable
or an RV Tauri variable, although GCVS classified it as a RV Tau of 67: day period with light 
amplitude of 2.3 magnitudes in visual. 
Earlier Harvard classifiers called it
SRd (Hoffleit 2000). Lloyd Evans (1999) has studied the total visual light
amplitude of RV Tauri stars with respect to spectral type across the 
instability strip. For stars of CE Vir spectral type (K2), the light amplitude
expected is about 0.3 not 2.3 magnitudes. The radial velocity variations
(shown in Table~1) do not suggest a large pulsation amplitude. However the 
 B-V colour of 1.39 noted at maximum light by Dawson \& Patterson (1982) is consistent
 with the colour expected of a K2 Supergiant (1.36). 
 
\begin{table}
\caption{Photometry of CE Vir}
\begin{tabular}{rllll}
\hline
     JD/date  &   V   & B-V & U-B &  Ref. \\
\hline \hline\\
  2442886.7  & 10.71& 0.99& ...  &        Dawson (1979)         \\
     2887.7 & 10.64& 0.98& ...  &                              \\
     4691.74  &  8.66& 1.39& 1.19 & Dawson and Patterson (1982) \\
     4704.71  &  9.13& 1.43& 1.46 &                             \\
              &      &     &      &                               \\
              &   J  &   H &  K   &                \\
              & 6.259   &  5.616 &   5.407 &  2MASS \\
              & 5.964   &   ...  &   4.594 &  DENIS \\ 
\hline
\multicolumn{5}{l}{Note:- J, H, and K magnitudes have typical errors of 0.02 (2MASS)}\\
\multicolumn{5}{l}{and 0.07 (DENIS) magnitudes.}
\end{tabular}
\end{table}

Dawson (1979) lists CE Vir as SRd with a period of 85 days.
There seems to be some dichotomy about the period and luminosity  if CE Vir is 
assumed to be a RV Tau star obeying the period luminosity relation (Pollard
\& Lloyd Evans 1999). The $T_{\rm eff}$ and log~$g$ obtained (see below) 
places the star on 
a log~L/L$_{\odot}$ $\approx$ 3.5 track for post-AGB stars passing through the RV Tauri 
instability strip in the  log~$g$ - $T_{\rm eff}$ plane
(Giridhar et al. 2000). This value of luminosity suggests a period of 330 days.
On the other hand if a period of 67 days is assumed the $M_{\rm v}$ expected is about
1.33 and is inconsistent with the luminosity estimated from $T_{\rm eff}$ and log~$g$.

\begin{figure*}
\epsfxsize=16truecm
\epsffile{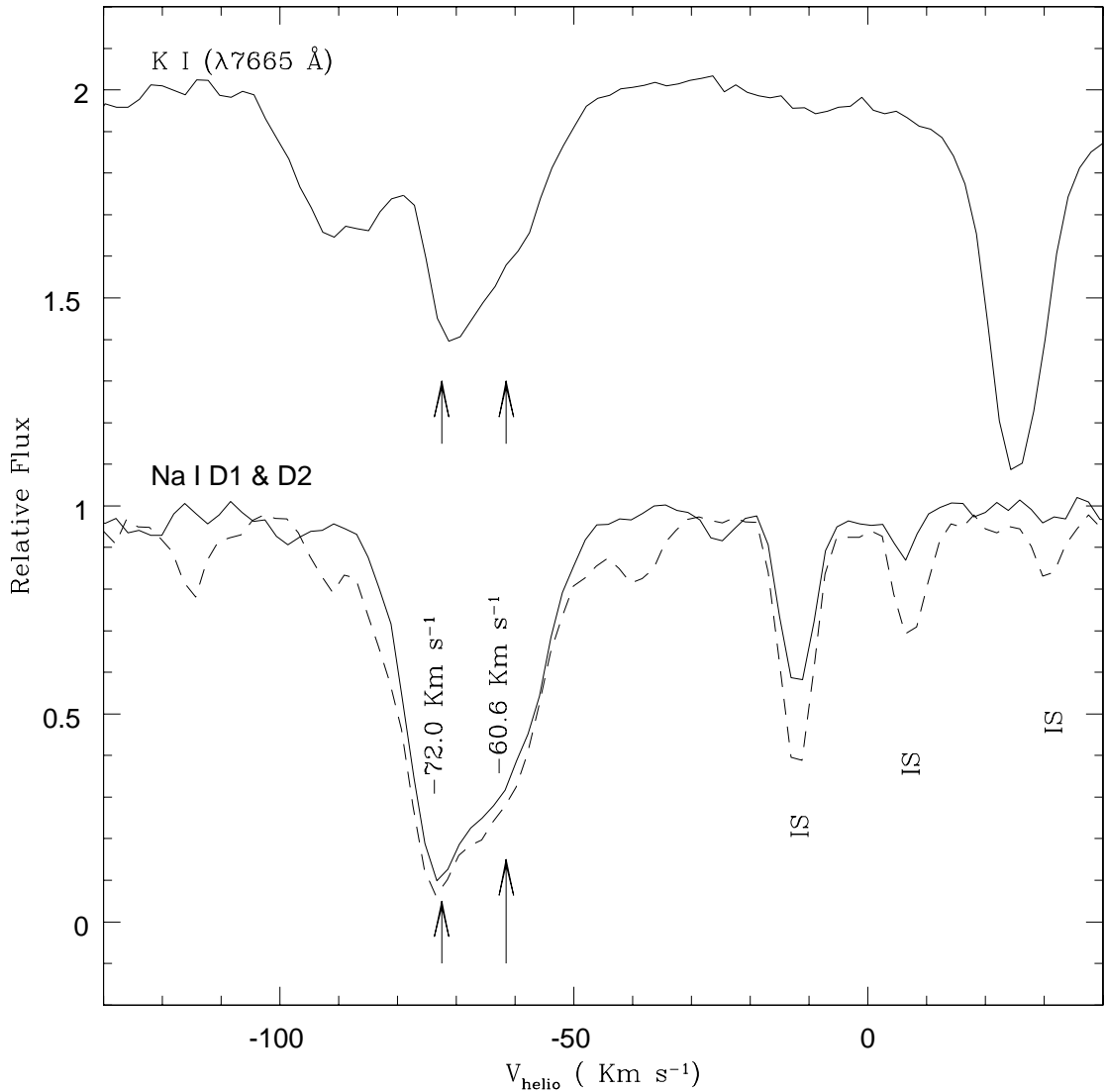}
\caption{ Profiles of Na D1 (solid) \& D2 (broken) and K~I ($\lambda$7665~\AA)
are shown in Heliocentric radial velocity units. Interstellar components
are marked as IS.}
\end{figure*}

\section{Description of the Spectrum}

Apart from the general weakness of the line spectrum, the spectrum of CE Vir
is similar to $\alpha$ Boo. The lines in CE Vir are slightly broader and thus
blended than in $\alpha$ Boo. A few aspects are different. As was discovered by
Gonzalez et al. (1997b) Li\,{\sc I} I $\lambda$ 6707 line is very strong in 
CE Vir. The Na I D lines are accompanied by 2 (or 3) sharp interstellar
components (Figure~2) that are red displaced relative to stellar lines and 
are present at radial velocities of
$-$11~km~s$^{-1}$, 8~km~s$^{-1}$ (and possibly at 32~km s$^{-1}$). 
The stronger component
at $-$11~km~s$^{-1}$ is also present in K\,{\sc I} $\lambda$ 7665 resonance line 

In addition to the interstellar components, both Na~I and K\,{\sc I} resonance
lines seem to have red displaced circumstellar components (Figure~2). The 
components are displaced by about 10~km s$^{-1}$ from the main stellar line.
If the stellar and the red circumstellar components are assumed to be gaussians
(mainly for K\,{\sc I} lines) the deblended equivalent widths of the components 
can be estimated.
We examined whether other resonance lines show such components. In our observes spectral
range Fe~I lines $\lambda$ 5110 and $\lambda$ 5060 show red absorption
components at the same displaced velocity. Similar red displaced 
components are also observed in the suspected RV Tauri star QY Sge (Rao, 
Goswami, \& Lambert 2003). This gas is either falling back after an expansion
due to pulsation or some infall from a circumstellar reservoir. Since only 
Na~I D, K\,{\sc I} and Fe~I resonance lines show these components, the gas must be
very cool and neutral.

As already mentioned H$\alpha$ shows a deep absorption 
corresponding to stellar velocity flanked by emission on each side symmetrically
displaced to either side by 32.5~km s$^{-1}$. The profile is also similar
on the two occasions observed  in 1996 by Gonzalez et al. (1997b).
Only the central absorption and emission peaks are slightly displaced
by few km s$^{-1}$.

\begin{figure*}
\epsfxsize=16truecm
\epsffile{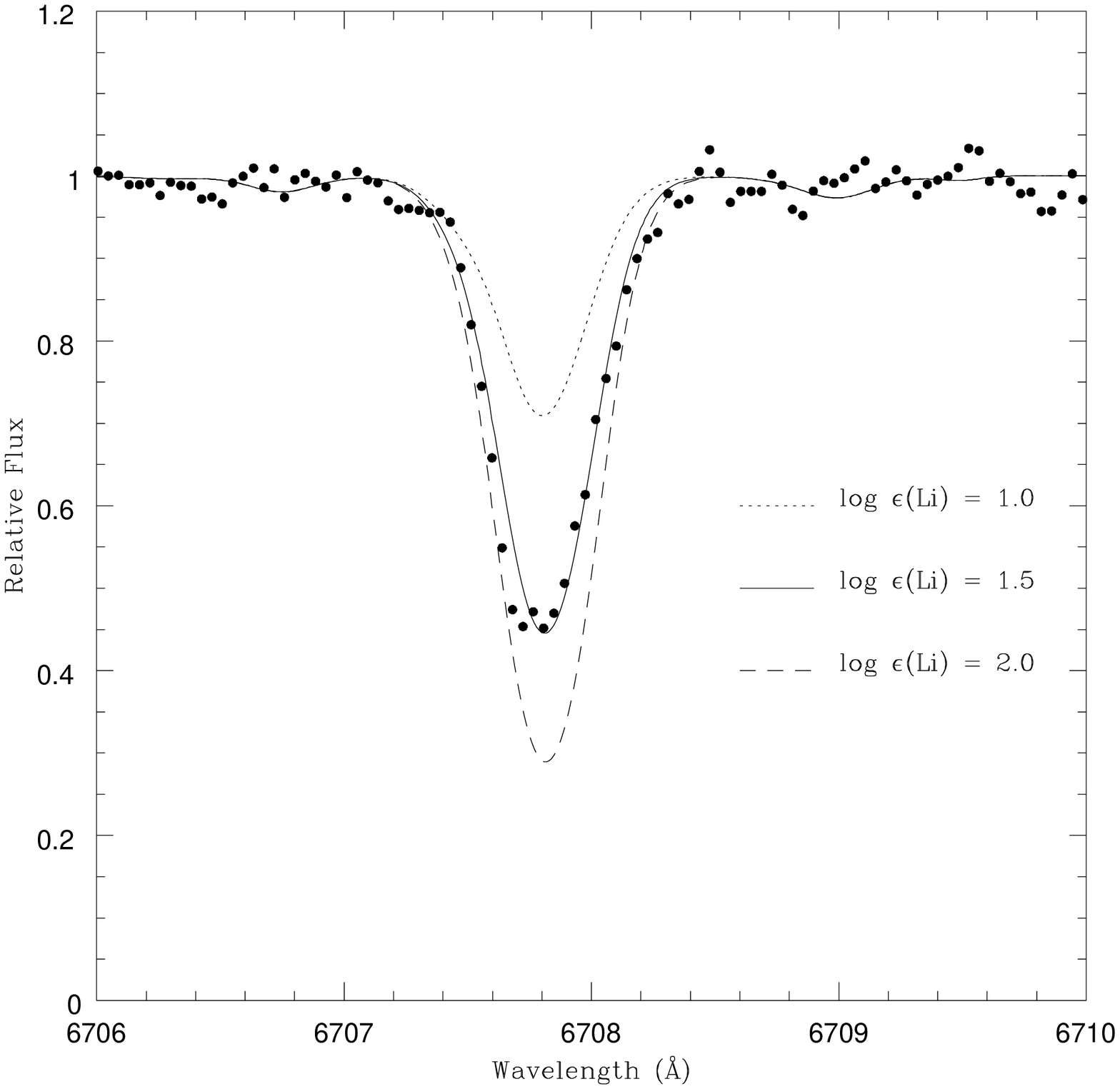}
\caption{Determination of total Li abundance from spectral synthesis of
Li~I line at 6707.6~\AA. Synthetic spectra (lines) have been computed for three different Li abundances
and compared with the observed Li profile (filled circles).} 
\end{figure*}

\begin{table}
\caption{Na~I and K\,{\sc I} IS components}
\begin{tabular}{lll}
\hline
              &  Rad.vel    & Eq.W  \\
              & km s$^{-1}$ & m\AA   \\
\hline\hline
{\underline{Comp.1}}      &             &          \\
Na~I D2     &  $-$11.3    &  92      \\
~~~~~~~D1     &  $-$11.35   &  64       \\
              &           &      \\
K\,{\sc II} ($\lambda$ 7665) &$-$11    &  17       \\
                  &       &        \\
{\underline{Comp.1}}      &             &          \\
Na~I D2       & 8.2     &  (*)     \\
~~~~~~~D1       & 7.3     &  12      \\
\hline
\multicolumn{3}{l}{* - contaminated by terrestrial H2O line.}
\end{tabular}
\end{table}

\begin{figure*}
\epsfxsize=16truecm
\epsffile{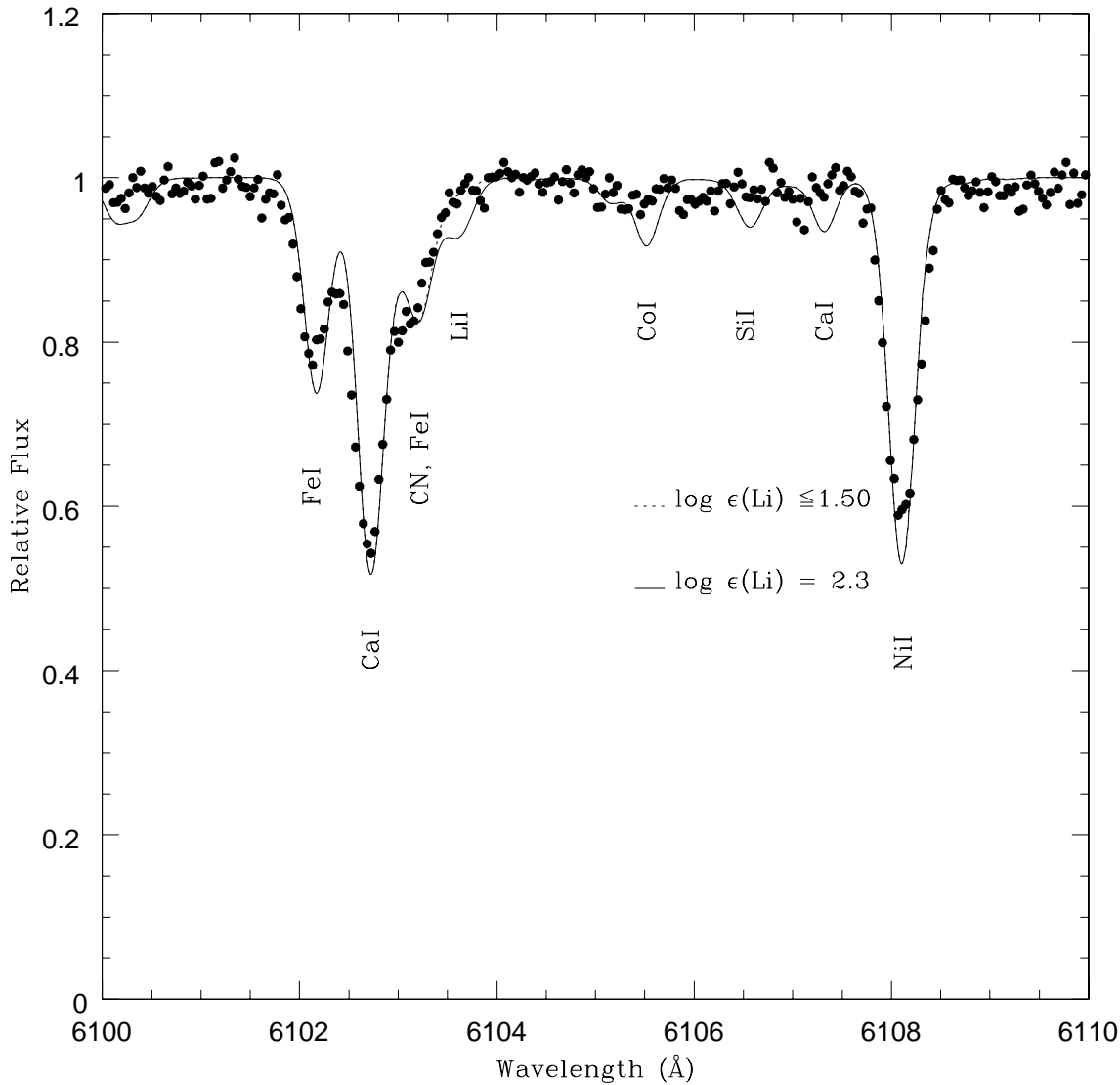}
\caption{Estimating  Li abundance by comparing predicted spectra (lines) for two different
Li abundances with the observed (filled circles) Li~I profile at 6103~\AA.}
\end{figure*}

\begin{figure*}
\epsfxsize=16truecm
\epsffile{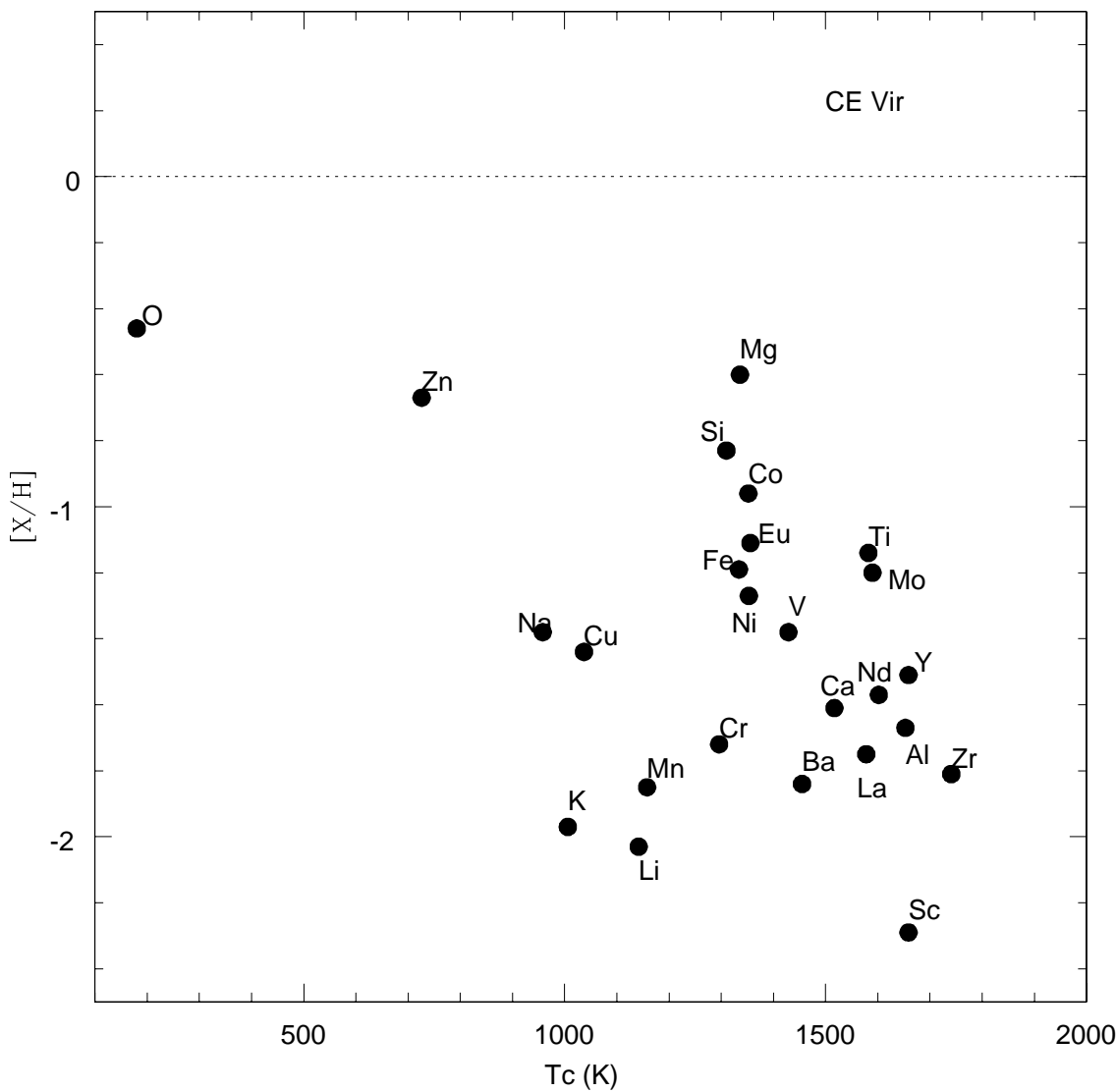}
\caption{Abundance depletions [X/H] are shown against condensation temperature.}
\end{figure*}

\begin{figure*}
\epsfxsize=16truecm
\epsffile{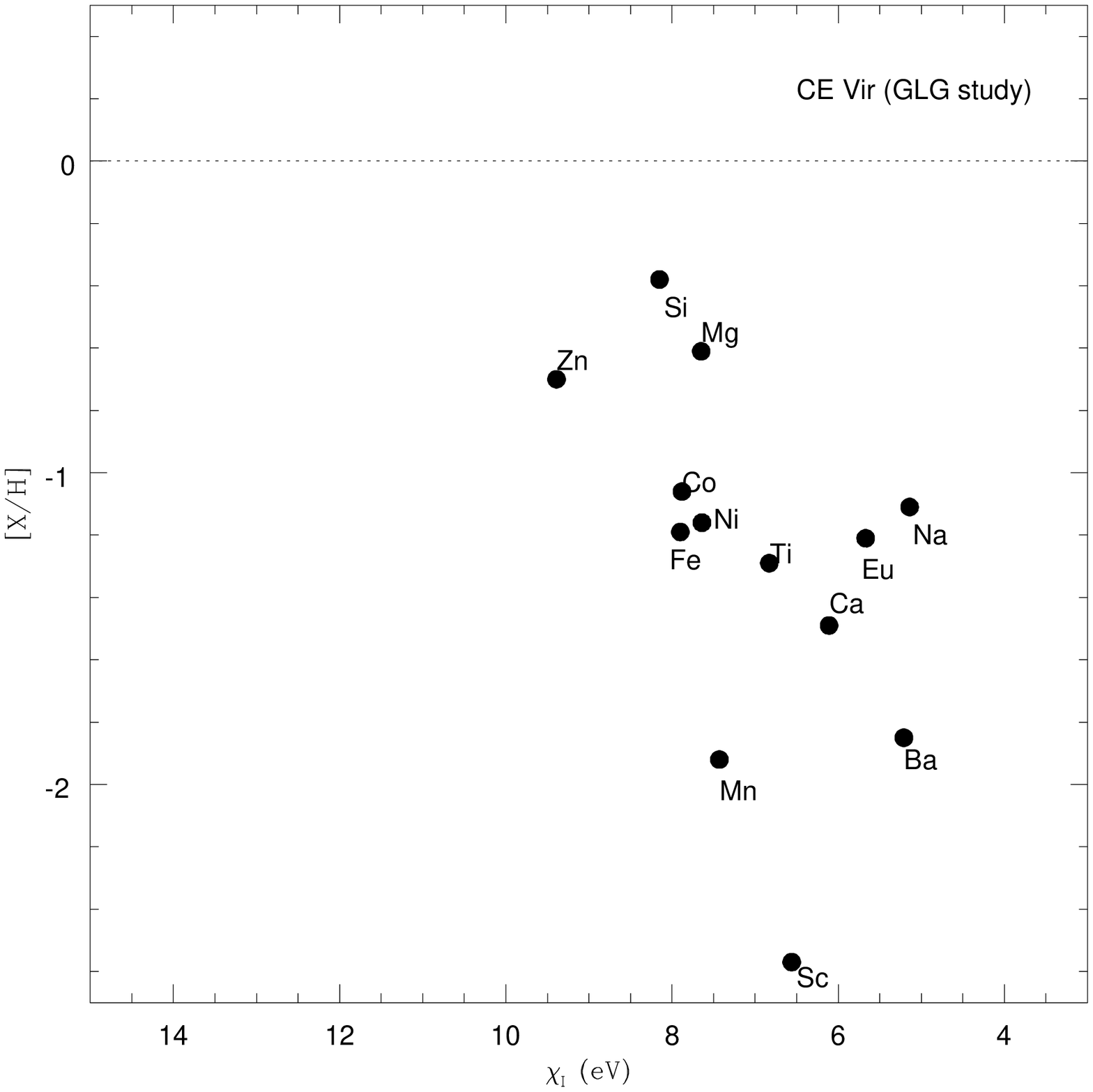}
\caption{Abundance depletions [X/H] (from GLG) are shown against first ionization potential.}
\end{figure*}

\begin{figure*}
\epsfxsize=16truecm
\epsffile{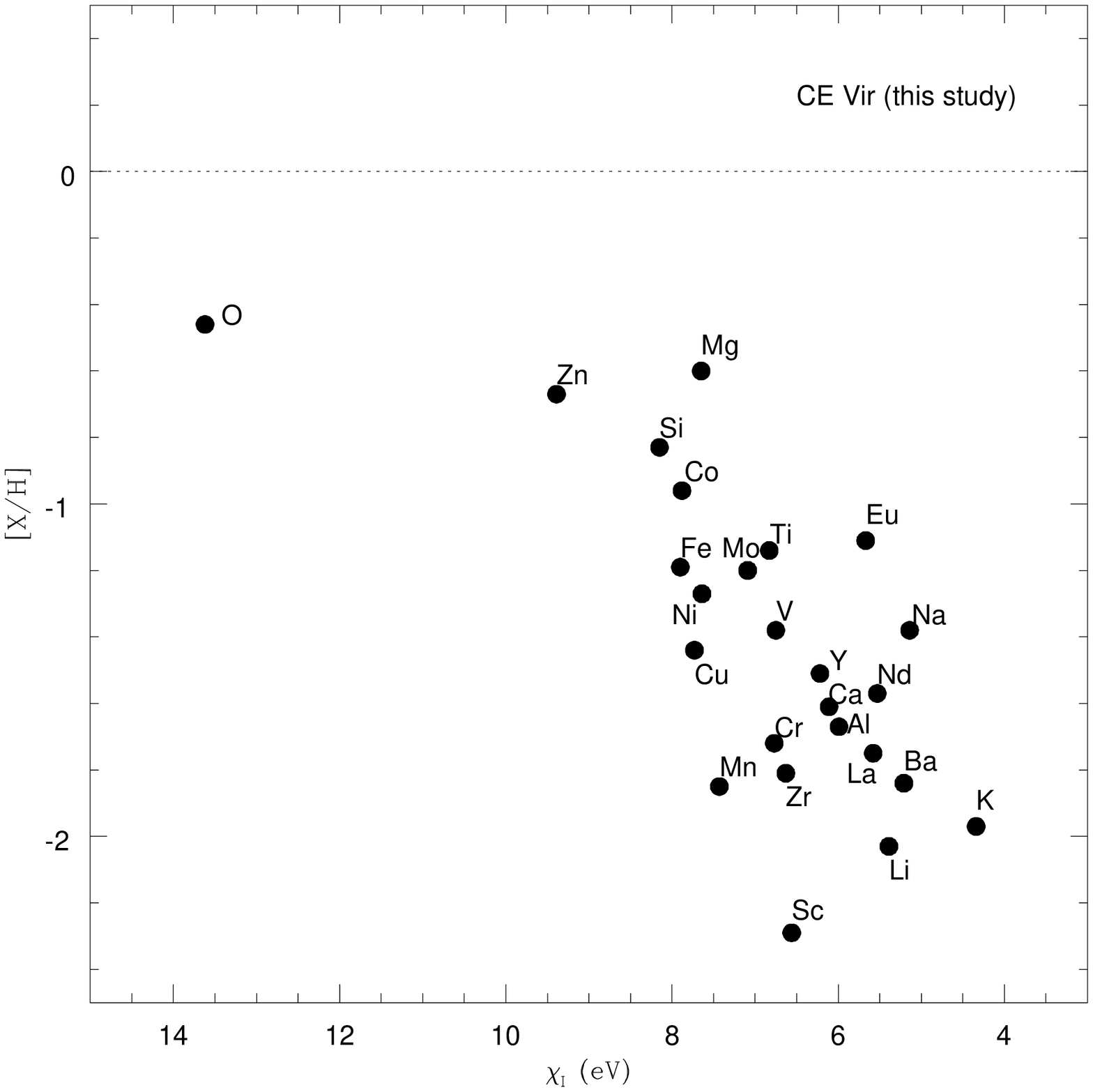}
\caption{Abundance depletions [X/H] are shown against first ionization potential.}
\end{figure*}

\begin{table}
\begin{center}
\caption{Abundance Summary for the CE VIR.
Solar abundances and the condensation temperatures ($T_{\rm C}$ for
the solar system composition
are adopted from (Lodders 2003). The last column $\chi_{\rm I}$ is the
first ionisation potential of the element X.}
\begin{tabular}{cc|cccccc}
\hline\hline
Species & log~$\epsilon$ (X$_{\sun}$)  & n  & log~$\epsilon$ (X) &  [X/H] & 
$T_{\rm C}$ & $\chi_{\rm I}$ \\
        &                              &     \multicolumn{3}{c}{4300/0.25/-1.0/3.40}& (K)& (eV) \\
\hline
Li\,{\sc I}  & 3.28 & 1 & 1.50 & $-$2.03 & 1142 & 5.39 \\
O\,{\sc I}   & 8.69 & 2 & 8.23 & $-$0.46 & 180  & 13.62  \\
Na\,{\sc I}  & 6.30 & 3 & 4.92 & $-$1.38 & 958  & 5.14 \\
Mg\,{\sc I}  & 7.55 & 6 & 6.95 & $-$0.60 & 1336 & 7.65 \\
Al\,{\sc I}  & 6.46 & 3 & 4.79 & $-$1.67 & 1653 & 5.99 \\
Si\,{\sc I}  & 7.54 & 5 & 6.71 & $-$0.83 & 1310 & 8.15 \\
K\,{\sc I}   & 5.11 & 2 & 3.17 & $-$1.94 & 1006 & 4.34 \\
Ca\,{\sc I}  & 6.34 & 8 & 4.73 & $-$1.61 & 1517 & 6.11 \\
Sc\,{\sc I}  & 3.07 & 1 & 0.78 & $-$2.29 & 1659 & 6.56 \\
Sc\,{\sc II} & 3.07 & 2 & 0.60 & $-$2.47 & ...  & ...\\
Ti\,{\sc I}  & 4.92 & 12& 3.78 & $-$1.14 & 1582 & 6.83\\
Ti\,{\sc II} & 4.92 & 5 & 4.01 & $-$0.92 & ...  & ... \\
V\,{\sc I}   & 4.00 & 4 & 2.62 & $-$1.38 & 1429 & 6.75 \\
V\,{\sc II}  & 4.00 & 2 & 2.86 & $-$1.14 & ...  & ... \\
Cr\,{\sc I}  & 5.65 & 5 & 3.93 & $-$1.72 & 1296 & 6.77 \\
Cr\,{\sc II} & 5.65 & 3 & 4.39 & $-$1.26 & ...  & ...\\
Mn\,{\sc I}  & 5.50 & 2 & 3.65 & $-$1.85 & 1158 & 7.43 \\
Fe\,{\sc I}  & 7.47 & 21& 6.28 & $-$1.19 & 1334 & 7.90 \\
Fe\,{\sc II} & 7.47 & 5 & 6.22 & $-$1.25 & ...  & ...    \\
Co\,{\sc I}  & 4.91 & 7 & 3.95 & $-$0.96 & 1352 & 7.88\\
Ni\,{\sc I}  & 6.22 & 12& 4.95 & $-$1.27 & 1353 & 7.64 \\
Cu\,{\sc I}  & 4.26 & 2 & 2.82 & $-$1.44 & 1037 & 7.73 \\
Zn\,{\sc I}  & 4.63 & 1 & 3.96 & $-$0.67 & 726  & 9.39 \\
Y\,{\sc I}   & 2.20 & 1 & 0.69 & $-$1.51 & 1659 & 6.22 \\
Zr\,{\sc I}  & 2.60 & 2 & 0.79 & $-$1.81 & 1741 & 6.63 \\
Zr\,{\sc II} & 2.60 & 1 & 1.28 & $\geq$$-$1.32 & ...  & ...  \\
Mo\,{\sc I}  & 1.96 & 2 & 0.76 & $-$1.20 & 1590 & 7.09 \\
Ba\,{\sc II} & 2.18 & 1 & 0.34 & $-$1.84 & 1455 & 5.21 \\
La\,{\sc II} & 1.18 & 2 & $-$0.57 & $-$1.75 & 1578 &  5.58\\
Nd\,{\sc II} & 1.46 & 1 & $-$0.11 & $-$1.57 & 1602 &  5.53\\
Eu\,{\sc II} & 0.52 & 1 & $-$0.60 & $-$1.11 &  1356 & 5.67\\
\hline
\end{tabular}
\end{center}
\end{table}
\section{Abundance analysis}

Gonzalez et al.'s (1997b) analysis was confined to a few elements.
In the present analysis, we tried to obtain estimates of the line strengths of as 
many elements as we can identify unambigously. We follow the
traditional LTE, Kurucz model atmospheres based analysis for a given
metallicity parameter([M/H]), using the current version of MOOG (Snedan 1973). One of 
the critical choice to be made is the assumed model [M/H] value. It is obvious
from Gongalez et al.'s (1997b) analysis that the Si and Mg have different
abundances (i.e [X/H]$\simeq$ $-$0.7 ) to Fe ( [Fe/H]$\simeq$ $-$1.3) -the main electron
donors to continuous opacity. A change in [M/H] from $-$0.7 to $-$1.3 changes the
abundance of ionised metals by 0.15~dex.  We have assumed [M/H] of $-$1.0
(same choice as Gonzalez et al. 1997b). This leads to an 
uncertainty  of around 0.07~dex. The $gf$ values have been obtained 
from the following sources: Reddy et al 2003, Lambert et al 1996, and compilation
by R. E. Luck (private communication).
The excitation and ionisation
equilibria are taken care by ensuring the same abundance for both high and low
excitation lines of Fe\,{\sc I} as well as Fe\,{\sc I} and Fe\,{\sc II}. The ionization equilibrium 
from Ti\,{\sc I}/ Ti\,{\sc II} and Cr\,{\sc I}/Cr\,{\sc II} are also kept satisfied. The final model
arrived at is $T_{\rm eff}$ of 4300~K, log~$g$ = 0.25, the microturbulent
velocity $\xi_{\rm t}$ = 3.4 km s$^{-1}$, and
[M/H] of $-$1.0., some what similar to the parameters obtained by Gonzalez et al.
(1997b). 
The resulting abundances are displayed in table 4. The Li abundance is 
arrived at by synthesizing the spectrum in $\lambda$ 6707 region (Figure~3) with
critically examined line list (Reddy et al. 2002) in the vicinity of the Li\,{\sc I} $\lambda$6707.
The resonance line of Li\,{\sc I} $\lambda$ 6707 is a blend of $^{6}$Li and $^{7}$Li isotopes and their 
multiple
HFS components. Wavelengths and log~$gf$ values for all the components were adopted from
Hobbs, Thorburn, \& Rebull (1999). In computing Li abundance, isotopic ratio $^{6}$Li/$^{7}$Li = 0.0
has been assumed. With the above input atomic data and the derived model, we obtained Li abundance
log~$\epsilon$(Li) = 1.5$\pm$0.2. 
The excited Li~I line at $\lambda$ 6103 is very weak and we could
only estimate an upper limit of 1.5~dex by way of spectrum synthesis (Figure~4),
using the line list from Kurucz (1994).
This is quite different
from the value $-$0.40 to $-$0.03 given by Gonzalez et al. (1997b).

An upper limit to carbon abundance estimated by synthesizing the C\,{\sc I} line
$\lambda$ 5380 (within the detection limit) is found to be $\leq$ 7.76. A
Nitrogen abundance of $\leq$ 7.0 is estimated using the upper limit of C and 
synthesizing the red CN lines in the $\lambda$ 8030 region. The oxygen abundance
is obtained from the strong [O I] lines 6300~\AA\  and 6363~\AA. Clearly C/O $<$ 1.

The K abundance is estimated from K~I resonance lines making an allowence for
the redward circumstellar component (an estimate is also made without making
this allowence) thus might be uncertain.

\section{Abundance Trends}

It has been suggested by Gonzalez et al. (1997b) that the 
pattern of abundances in CE Vir are generally like that seen in warm RV Tau
stars except that only the elements with highest $T_{\rm c}$ are depleted.   
It is obvious from their figure that the abundances of Na and Mn are not consistent with that suggestion.
Both are much more depleted than Si and Mg which have much higher $T_{\rm c}$. In the
entire sample of RV Tau stars analysed Na is either enhanced or of the same
depletion value as Zn, S., never much less -thus CE Vir is unusual.
We compared our values of [X/H] with those obtained by Gonzalez et al.
They show no major differences except for Li. 
The average difference for 13 elements (ours - GLG) is just $-$0.007$\pm0.11$.
The depletion
of Na and Mn are confirmed (different lines used than those by GLG).  
 
In Figure~5, we plot the $T$$_{\rm c}$ versus our  [X/H] from Table~4. $T_{\rm c}$ here is taken
as the condensation temperature where 50$\%$ gas is condensed into solid form
as estimated for solar system abundances and a pressure of 10 $^{-4}$ bar
by  Lodders (2003). It is obvious from the figure that no single systematic pattern
emerges. This  is totally untypical of warm RV Tau stars (eg. IW Car - 
Giridhar, Rao \& Lambert 1994). It is apparent that the abundance pattern 
exhibited by CE Vir is neither consistent with that generally expected from metal
poor stars nor warm RV Tau stars.
  
On the other hand in Figure~6 the [X/H] values obtained by GLG are plotted
with respect to the first ionization potential (FIP) of the element. Figure shows 
a definite trend, although some scatter is present; the lower the 
ionization potential the higher is the depletion.  Our more extensive data
from Table~4 are plotted in the Figure~7. Clearly it is obvious from the Figure~7,
that for element 
whose first ionization potential is below 8~eV depletion occurs, lower the
ionization potential higher is the depletion. Even Na shares this pattern.
Sc does seem to be slightly more depleted than the trend for other elements.

Giridhar, Lambert \& Gonzalez (1998, paper IV) did examine the possibility
of depletions being correlated with first ionization potential of the element
in other warmer RV Tau stars eg. AD Aql. They however discounted this 
possibility on the grounds that the alkalis Na and K and Al do not fit into the
ionization potential versus [X/H] trend of other elements. Secondly the
 trend with $T_{\rm c}$ has less scatter. The situation in CE Vir is quite different.
All three so called discordent elements do fit into general trend. K 
abundance is estimated from two resonance lines, which have red displaced
circumstellar components. (the abundance or deficiency  estimate is not
affected  much by either including or excluding the circumstellar component).
Even Li fits this pattern quite well (coincidence?).
If real, this  implies that the undepleted
abundance of Li is about solar system value of 3.31 which suggests Li has been re-manufactured
during its evolution.

\section{Discussion} 

Abundance correlations with ionization potential in Pop~II variables has been in vogue 
for a long time eg. W Vir (see the foot note in Barker et al. 1971). In CE Vir
this clearly seems to be present for elements with first ionization potentials
below $\simeq 8~eV$. It is likely that singly ionized elements escaped as 
stellar wind probably controlled by magnetic fields (open field lines ?) 
rather than coupled to radiation pressure on dust (presence of dust in the
atmosphere is not convincing in this and other similar stars e.g.,DY Aql) .
  
What provides the extra source of ionization to the atmospheric gas is not 
clear. The velocity variations, the amplitude of light curve and the period of
pulsation are very uncertain presently for CE Vir. The pulsation amplitude
is likely to be small ( based on the radial velocity measurements obtained so
far). Whether atmospheric shocks can provide this ionization is not clear. It is true
that the star shows H$\alpha$ emission like a Be star - a disk or shell of 
emitting gas is present. The photon energies needed to ionize H would be 
too high for the trend seen here.

Is there still some evidence for dust-gas separation ($T_{\rm c}$ versus abundance 
depletion$?$) Sc abundance does suggest in a mild way such a possibility 
considering the extra depletion suffered.  Does circumstellar dust exist in
the environment of CE Vir? Recently (since the paper by GLG 2000) the 2MASS (Cutri et al. 2003) 
and DENIS
near IR photometry became available and is shown in Table~2.

\begin{figure*}
\epsfxsize=16truecm
\epsffile{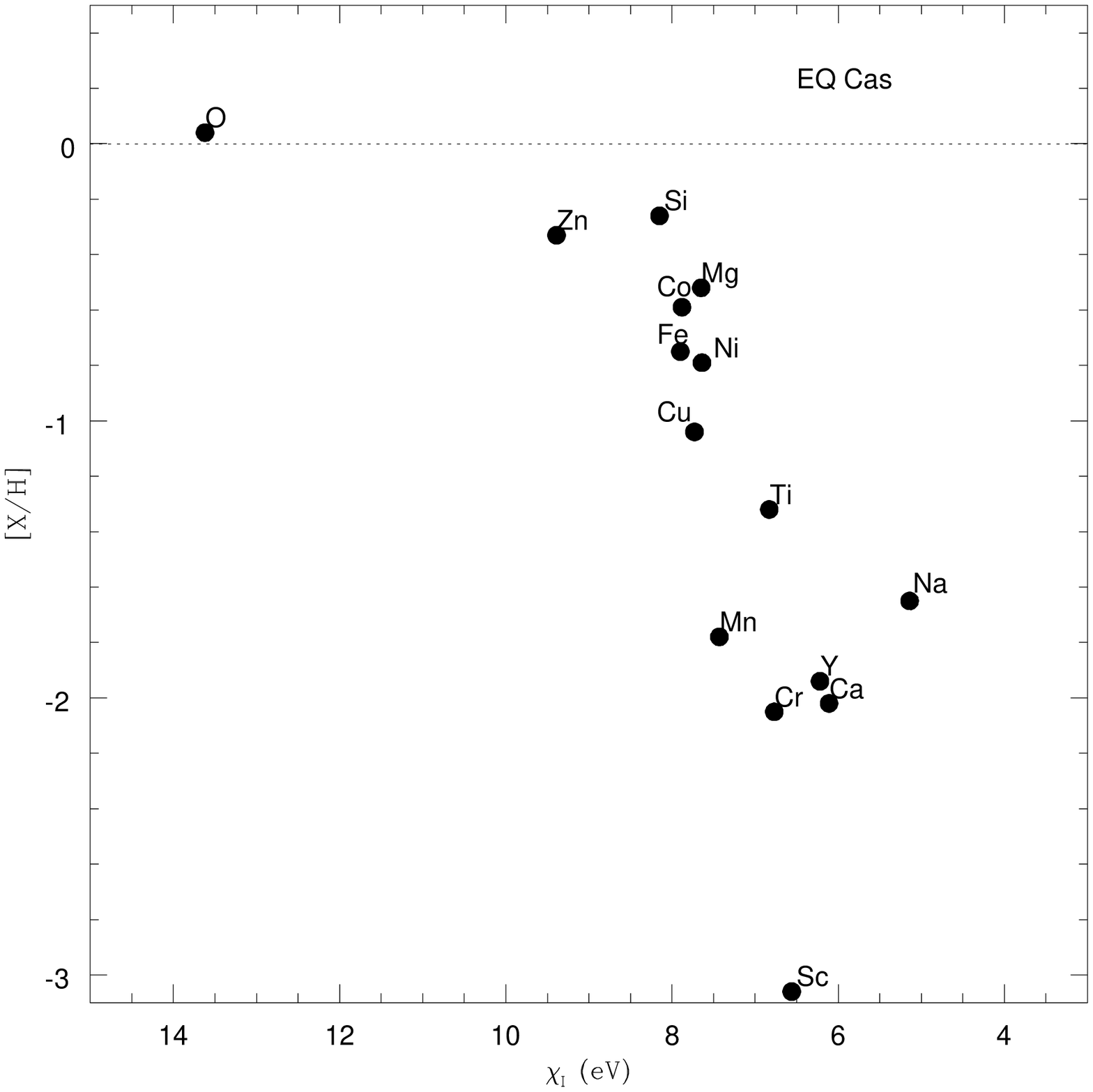}
\caption{ Abundance depletions [X/H] against first ionization potential
for EQ Cas. Abundance results are taken from Giridhar et al. (2004).}
\end{figure*}

The two sets of measurements do show variability of the near IR flux. The
J magnitude changed by 0.3, more importantly the K magnitude changed by
0.8. The near IR colours of RV Tau stars have been discussed by Goldsmith
et al. (1987). 2MASS colours of CE Vir are marginally outside the region seen
for G to M giants in J-H versus H-K diagram (other cool RV Tau star, DY Aql lies
within this region). But DENIS colours are appreciably different and suggests
excess radiation particularly at K band, indicative of the presence of hot dust.
The variability of K magnitude  probably suggests sporadic formation of dust. 
In summary there is evidence for presence of hot ($\simeq 1200 - 1300$ K )
dust.

The picture that emerges is that stellar wind, probably controlled by 
magnetic fields, (see Pascoli 1997; Garcia-Segura et al. 2004
for magnetic field driven winds in post-AGB stars) puts 
the gas into circumstellar regions in which sporadic
dust formation occurs. There is also evidence from resonance lines of Fe\,{\sc I},
K~I and Na~I that some of the cool gas (coinhabited with dust ?) returns
to the star. 

CE Vir, being the coolest member of the group, might show the initiation
of the dust-gas separation activity.  It is unlikely that cool stars like
DY Aql, CE Vir have enormous amounts of dust manufactured in their atmospheres
which would then drag the gas away by radiation pressure. If dust is 
manufactured in some cool circumstellar region (disk, shell), radiation 
pressure on dust can not be the source to put the photospheric gas into
the shell. The other likely mechanism is the pulsation. It is not clear atleast in
CE Vir, the pulsation would be strong enough to eject gas (and start 
stellar wind). If it is not strong enough, the other mechanism could be the
magnetic field driven wind pulling the ionized gas with it. CE Vir suggests
such a possibility. Would such evolved stars still retain surface magnetic
fields is a question to be explored. What provides the extra ionization in
CE Vir is still a mystery.
         
Very recently Giridhar et al. (2004) further estimated surface 
abundances for a sample of dozen new RV Tauri stars. One of the stars, EQ Cas, is
unique and displays an  abundance pattern that is quite different from other 
RV Tauri stars that were affected by dust - gas separation. However it shows a
great similarity to the abundance pattern presented in this paper for CE Vir.  
Elements like Na, Mn etc. whose $T_{\rm c}$s are much lower than that of Si, Mg etc.
are much more depleted than Si and Mg, contrary to what is expected in dust
condensation scenario.  However the same depletion values ([X/H]) are well
correlated with first ionization potentials for elements below 8~eV (Figure 8),
as seen for CE Vir. Thus EQ Cas looks like a twin of CE Vir. EQ Cas also shows
at certain phases  $T_{\rm eff}$ of 4500~K, log~$g$ = 0.0,  micro
turbulent velocity $\xi_{\rm t}$ = 4.6~km s$^{-1}$, and
[M/H] of $-$0.8 similar to CE Vir. It is also a high velocity star with a
radial velocity of $-$158 km s$^{-1}$. 2 MASS data provides J, H, K magnitudes
for EQ Cas. The J - H , and H - K colours do not suggest near IR flux excess.

The phenomenon shown by CE Vir and EQ Cas might not be uncommon (2 out
of about 35 stars) among RV Tauri stars. Particularly cooler stars might be
more prone.  
The abundance patterns displayed by both CE Vir and EQ Cas, namely
a  strong correlation of increasing elemental depletion with decreasing FIP,
suggest operation of stellar wind that selectively removed the 
low FIP elements from the photospheres of the stars. A striking similarity
can thus be seen in composition with slow solar wind. It is long
been known that slow solar wind, which arises mainly from lower solar latitudes,
and the solar corona both show a dependance of elemental abundances on 
the first ionization potential. The 
elements with FIP $<$ 10~eV (eg., Mg, Si, Fe etc ) show abundance enhancements
by a factor of 4 relative to their photospheric values, where as the
elements with FIP $>$ 10~eV (eg. O, S, Ne etc.,) do not show such 
enhancements - the FIP effect (Geiss 1982; Geiss et al. 1995). Although there is no unique
model that explains this elemental fractionation, it is believed that 
magnetic fields play a vital role (Schwadron et al 1999, Laming 2004).
However, such a fractionation is thought to be in the upper 
chromosphere (Henoux 1998) where the low FIP elements get ionized (by UV
radiation ) but not the high FIP elements and the particle density is not
high enough for collisions to couple ionized and neutral atoms. Neutral
high FIP elements can escape from magnetic structures perpendicular to the
lines of force where as the ionized low FIP elements are confined. Densities
where magnetic field is strong enough to satisfy this condition
occurs mainly in the chromosphere.  FIP effects are also seemed to be 
present in stellar coronae of solar type (inactive) stars (eg. $\alpha$
Cen ) where as some
active stars (eg. RS CVn stars)show even inverse FIP effect (see Linsky 2002, Drake
2003 
for a review of stellar coronae). Even single late type giants show solar
like FIP effect, despite their clear non-solar evolution and internal
structure (Garcia-Alvarez et al 2004).

Although Sun and other solar type stars show the FIP effect in their 
coronae (and possibly winds) their photospheres are unaffected. In case of
CE Vir and EQ Cas the photospheres show the leftover gas after 
FIP effected wind has modified the atmosphere. It is possible either that
the elemental fractionation could have occured (or occuring) on the
photospheres or fractionated gas has been accreted on to the photosphere
from circumstellar regions that were swept by FIP effected wind.   
Thus, the realisation of counter FIP effect (deficiency correlation with FIP) in CE Vir and EQ Cas
aquire a special significance.
It would be of great interest to trace in CE Vir and EQ Cas the presence 
of either the stellar 
wind, its composition and/or chromosphere or corona that might show
FIP effect. 

\section{Acknowledgements}
We would like to thank Guillermo Gonzalez for sending us 1996 spectra 
of CE Vir and his comments. We also would like to thank Gajendra Pandey and Aruna Goswami for
their comraderie while observing at VBT. We would like to express our 
appreciation to S.Sriram, F.Gabriel, K.Jayakumar and Anbalagan for the upkeep
of the echelle spectrometer at VBT. 
We thank the anonymous refree for 
his comments and especially suggesting the relavence
of FIP effect in Sun and other stellar coronae.  
We would like to acknowledge with thanks
use of 2MASS and DENIS data bases. Our thanks are also due to David Lambert for his
encouragement.

\label{lastpage}
\end{document}